# Predição da Idade Cerebral a partir de Imagens de Ressonância Magnética utilizando Redes Neurais Convolucionais

Victor H. R. Oliveira[1], Augusto Antunes[3], Alexandre S. Soares[1], Arthur D. Reys[2], Robson Z. Júnior[2], Saulo D. S. Pedro[2], Danilo Silva[1]

Para a Alzheimer's Disease Neuroimaging Initiative

e o Australian Imaging Biomarkers and Lifestyle flagship study of ageing

[1]Universidade Federal de Santa Catarina, Florianópolis, SC
[2]Grupo 3778, Belo Horizonte, MG
[3]Núcleo de Ensino, Pesquisa e Ensino Alliar - NEPIA, Belo Horizonte, MG

victoroliveira.eng@hotmail.com, augusto.antunes@alliar.com, solisoareswa@gmail.com, arthur.reys@3778.care, robson.zagre@3778.care, saulo.pedro@3778.care, danilo.silva@ufsc.br

***Abstract.*** *In this work, deep learning techniques for brain age prediction from magnetic resonance images are investigated, aiming to assist in the identification of biomarkers of the natural aging process. The identification of biomarkers is useful for detecting an early-stage neurodegenerative process, as well as for predicting age-related or non-age-related cognitive decline. Two techniques are implemented and compared in this work: a 3D Convolutional Neural Network applied to the volumetric image and a 2D Convolutional Neural Network applied to slices from the axial plane, with subsequent fusion of individual predictions. The best result was obtained by the 2D model, which achieved a mean absolute error of 3.83 years.*

***Resumo.*** *Neste trabalho são investigadas técnicas de aprendizado profundo para a predição da idade cerebral a partir de imagens de ressonância magnética, visando auxiliar na identificação de biomarcadores do processo natural de envelhecimento. A identificação de biomarcadores é útil para a detecção de um processo neurodegenerativo em estágio inicial, além de possibilitar prever um declínio cognitivo relacionado ou não à idade. Duas técnicas são implementadas e comparadas neste trabalho: uma Rede Neural Convolucional 3D aplicada na imagem volumétrica e uma Rede Neural Convolucional 2D aplicada a fatias do plano axial, com posterior fusão das predições individuais. O melhor resultado foi obtido pelo modelo 2D, que alcançou um erro médio absoluto de 3.83 anos.*

**Palavras-chave:** Aprendizado Profundo; Idade Cerebral; Imagem de Ressonância Magnética.

## INTRODUÇÃO

A identificação de biomarcadores no processo de envelhecimento natural do cérebro permite o avanço de estudos relacionados a esta área, auxiliando na compreensão sobre a relação entre a idade e o estado cognitivo do paciente, bem como o estudo de eventuais processos neurodegenerativos [1]. Um biomarcador indicativo da idade cerebral que esteja muito distante da idade cronológica de uma pessoa pode indicar problemas que se relacionam ao risco de comprometimento cognitivo e doenças neurodegenerativas [2].

Neste trabalho desenvolvemos um modelo de redes neurais convolucionais para predição da idade cronológica de pacientes cognitivamente saudáveis a partir de imagens estruturais do cérebro obtidas através de ressonância magnética, visando seu uso posterior como um biomarcador do processo natural de envelhecimento. Foram investigadas arquiteturas de Redes Convolucionais 2D e 3D, buscando obter o melhor desempenho com o menor custo computacional possível.

## MATERIAIS E MÉTODOS

### Conjunto de dados

Utilizamos as bases ADNI (Alzheimer's Disease Neuroimaging Initiative) e AIBL (The Australian Imaging, Biomarkers & Lifestyle Flagship Study of Ageing) (Figura 1). O conjunto ADNI consiste em um coorte com 2720 indivíduos adultos com idade entre 55-96 anos (idade média = 76,44, desvio padrão = 6,64) e o conjunto AIBL contém 724 indivíduos adultos com idade entre 60-92 anos (idade média = 73.09, desvio padrão = 6.37). Foram utilizados apenas os pacientes diagnosticados como saudáveis e livres de qualquer indício de problemas cognitivos (conforme http://adni.loni.usc.edu/study-design). Os dados foram adquiridos em aparelhos de ressonância magnética de alto campo (1.5T e 3.0T), utilizando a sequência T1-MPRAGE. Utilizou-se 80% do ADNI para treinamento e 20% para validação. Todo o conjunto de dados AIBL foi utilizado para testes.

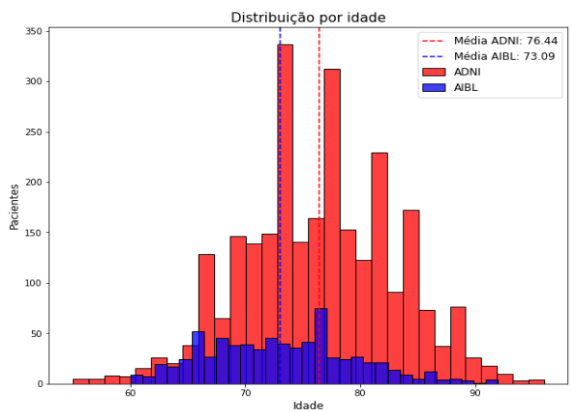

**Figura 1. Histograma de idades**

**Pré-Processamento**

Todas as imagens passaram por um *pipeline* de pré-processamento utilizando ferramentas do *software* FSL (https://fsl.fmrib.ox.ac.uk/fsl/fslwiki). Primeiramente foi aplicada a técnica de *skull-stripping*, que consiste no processo de segmentação do tecido cerebral, removendo tecidos de áreas não cerebrais. Posteriormente foi feito o co-registro, onde as imagens passam por uma normalização no espaço MNI152 (Figura 2), para alinhá-las a coordenadas padrão, garantindo a padronização quanto à orientação e a dimensão dos *voxels* em (2mm$^3$, 91 x 109 x 91).

Pretendendo eliminar a variação da intensidade de voxels entre as imagens, aplicou-se a técnica de *histogram matching*, permitindo que todas as imagens estejam equalizadas em relação a um histograma referencial obtido a partir de 50 volumes do ADNI. A normalização Min-Max foi aplicada em todos os volumes, transformando os valores em uma escala de 0-255.

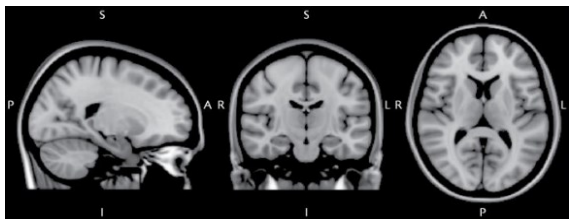

**Figura 2. Exemplo de uma imagem de Ressonância Magnética Anatômica no espaço MNI152, contendo os cortes sagital, coronal e axial, representados da esquerda para a direita.**

**Rede Convolucional 3D**

Foi desenvolvido um modelo de Rede Convolucional 3D, contendo uma arquitetura que consiste em 4 blocos. Cada bloco contém 2 camadas convolucionais, a função de ativação ReLU e uma camada de *Max Pooling*. Os filtros das camadas convolucionais iniciam-se em 16 no primeiro bloco e vão dobrando de tamanho a cada bloco, finalizando no tamanho 128. Nas camadas finais é aplicado um *Global Average Pooling* juntamente com uma camada densa com tamanho de 128 unidades, função de ativação ReLU e uma camada de *dropout* com valor 0.5. A última camada é densa, de saída linear, com apenas uma unidade.

Utilizou-se o otimizador *Adam*, com taxa de aprendizado inicial de $10^{-3}$, aplicando a técnica de decaimento linear por 80 épocas. Além do otimizador, aplicou-se técnicas de *data augmentation* que consistem em variações de brilho e contraste, flip horizontal e vertical, rotação em 10 graus, zoom e translação, sempre com pequenos valores para não criar imagens fora da distribuição do conjunto de dados. Como função de perda foi utilizado o erro médio quadrático (MSE). O tamanho do *batch* foi definido com o valor de 10.

**Rede Convolucional 2D**

A motivação para a escolha de um modelo 2D deve-se à utilização de técnicas de aprendizado por transferência utilizando um modelo pré-treinado na ImageNet [3]. Para cada volume (imagem 3D) foram escolhidas 40 fatias (imagens 2D) centrais do eixo axial. As fatias foram recortadas para eliminar regiões de pixels sem cérebro, resultando em dimensões de (86 x 104).

Durante o treinamento, cada fatia é considerada uma amostra independente, conforme [4]. Na etapa de inferência, as 40 fatias de um paciente são inseridas no modelo treinado e a mediana das predições individuais é utilizada como predição da idade do paciente.

O modelo foi desenvolvido utilizando a arquitetura da EfficientNetB3 [5]. A escolha dessa arquitetura foi feita devido a menor complexidade e melhor desempenho quando comparada a outras redes, como demonstrado nos testes realizados por [5]. De fato, em nossos testes preliminares comparando o desempenho da EfficientNetB3 com a inception-resnet-v2 utilizada em [4], o modelo da EfficientNetB3 demonstrou melhor desempenho em menor tempo de treinamento.

Para treinar o modelo, utilizou-se pesos pré-treinados da ImageNet. Nas camadas finais é aplicado um *Global Average Pooling* e uma camada densa com tamanho de 1024 unidades, seguida por uma camada de *dropout*, configurada com o valor de 0.5, com o objetivo de evitar o *overfitting*.

Utilizou-se o otimizador *Adam* com taxa de aprendizado inicial de $10^{-4}$, mantendo um decaimento linear por 50 épocas. O tamanho do batch foi definido como 32. O erro médio quadrático (MSE) foi utilizado como perda.

# RESULTADOS

A implementação foi realizada usando as bibliotecas TensorFlow e Keras, com recursos computacionais de uma GPU NVIDIA Tesla T4 x1. O treinamento do modelo 2D durou aproximadamente 180 segundos/época, enquanto o modelo 3D levou aproximadamente 117 segundos/época.

O erro médio absoluto (MAE) foi utilizado como métrica de avaliação. Os resultados das predições no conjunto de teste podem ser visualizados na Tabela 1.

**Tabela 1 – Resultados**

| Modelo | MAE (anos) |
|---|---|
| Rede Convolucional 3D | 4.66 |
| Rede Convolucional 2D | 3.83 |

A avaliação dos resultados indica um bom desempenho para o modelo 2D. A utilização de um modelo robusto de redes convolucionais 2D, inicializado com pesos pré-treinados na ImageNet, permitiu obter o melhor desempenho. Por meio da Figura 3, é possível analisar o resultado da inferência deste modelo sob o conjunto de testes.

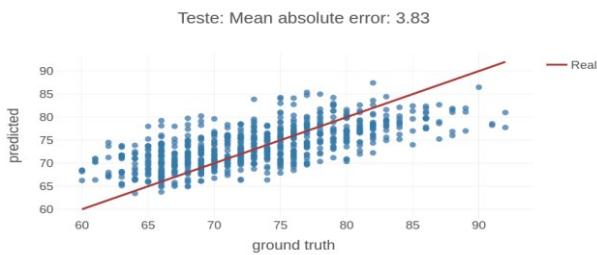

**Figura 3. Resultados do modelo 2D**

O atual estado da arte utilizando redes 2D [4] faz uso da rede inception-resnet-v2 treinada em um conjunto de dados com 11729 pacientes e obteve um MAE de 3.7 anos.

# CONCLUSÕES

Neste trabalho utilizou-se redes convolucionais 2D e 3D para predição da idade cerebral por meio de imagens de ressonância magnética. Os dois modelos foram treinados utilizando o conjunto de dados ADNI e testados no AIBL. O modelo 2D demonstrou o melhor desempenho, obtendo um MAE de 3.83 anos no conjunto de testes.

Para que o desempenho se aproxime ainda mais do estado da arte, alguns pontos ainda podem ser aperfeiçoados, como adicionar outros conjuntos de dados, envolvendo diferentes *scanners*, localizações geográficas e características de pacientes. Pretende-se também investigar diferentes abordagens de fusão das predições de fatias individuais produzidas pelo modelo 2D. Uma etapa posterior do projeto envolve a aplicação do modelo em imagens de sujeitos com comprometimento cognitivo, visando avaliar a diferença entre as idades cronológica e predita pelo modelo.